\documentclass[11pt]{article}
\usepackage{epsf}
\newtheorem{problem}{Problem}

\newcommand{\INPUT}[0]{{\bf Input}} 
\newcommand{\OUTPUT}[0]{{\bf Output}}

\newcommand{\MM}[0]{{\cal M}}
\newcommand{\BB}[0]{{\cal B}}

\newtheorem{lemma}{Lemma}[section]
\newtheorem{theorem}{Theorem}[section]

\def\Proof{\par\noindent{\sl Proof\/}:\enspace}
\def\blackslug{\hbox{\hskip 1pt \vrule width 4pt height 8pt depth 1.5pt
  \hskip 1pt}}
\def\QED{\quad\blackslug\lower 8.5pt\null}

\begin{document}

\title{Designing Proxies for Stock Market Indices \\ is
Computationally Hard\thanks{An abstract appeared in the Proceedings of
the 10th Annual ACM-SIAM Symposium on Discrete Algorithms, 1999.}}

\author{Ming-Yang Kao\thanks{Supported in part by NSF Grant
CCR-9531028.}\\Department of Computer Science\\
Yale University\\New Haven, CT\ \ 06520 
\and
Stephen R. Tate\thanks{Supported in part by NSF Grant
CCR-9409945 and Texas Advanced Research Program 
Grant 1997-003594-019.}\\Department of Computer Science\\University of North
Texas\\ Denton, TX\ \ 76203} 
\date{}

\maketitle

\begin{abstract}
In this paper, we study the problem of designing proxies (or
portfolios) for various stock market indices based on historical data.
We use four different methods for computing market indices, all of
which are formulas used in actual stock market analysis.  For each
index, we consider three criteria for designing the proxy: the proxy
must either track the market index, outperform the market index, or
perform within a margin of error of the index while maintaining a low
volatility.  In eleven of the twelve cases (all combinations of four
indices with three criteria except the problem of sacrificing return
for less volatility using the price-relative index) we show that the
problem is NP-hard, and hence most likely intractable.
\end{abstract}

\section{Introduction}
\label{sec:intro}

Market indices are widely used to track the performance of stocks or
to design investment portfolios \cite{asb93}.  This paper initiates a
rigorous mathematical study of the computational complexity of the art
of designing proxies for such indices.  There are several results on
selecting such proxies (or portfolios) in an on-line manner (see, for
example,~\cite{cover:91} and~\cite{co:96}), but we look at off-line
algorithms for designing proxies based on historical data.  In
particular, we show that, with one exception, all combinations of
three fundamental problems (such as tracking or outperforming a full
market index) with four commonly-used indices give NP-complete
problems, so are computationally hard.  We conjecture that the one
remaining problem is also NP-complete, but do not have a proof at this
time.

To formally define market indices, let $\BB$ be a set of $b$ stocks in
a market.  Let $S_{i,t}\geq 0$ be the price of the $i$-th stock at time
$t$.  Let $w_i$ be the number of outstanding shares of the $i$-th
stock.  We assume that $w_i$ does not change with time.  This paper
discusses computational complexity issues regarding four kinds of
market indices currently in use \cite{asb93}.  These indices are
calculated by the following formulas, which can be multiplied by
arbitrary constants to arrive at desired starting index values at time
0.

$\bullet$ The {\em price-weighted index} of $\BB$ at time $t$ is
\begin{equation}
\Phi_1(\BB,t)=\frac{\sum_{i=1}^b{S_{i,t}}}{b}.
\label{eq:phi1def}
\end{equation}
The Dow Jones Industrial Average is calculated in this manner for some
$\BB$ consisting of thirty stocks.

$\bullet$ The {\em value-weighted index} of $\BB$ at time $t$ is
\[\Phi_2(\BB,t)=\frac{\sum_{i=1}^b{w_i{\cdot}S_{i,t}}}{\sum_{i=1}^b{w_i{\cdot}S_{i,0}}}.\]
The Standard \& Poor's 500 is computed in this way with respect to 500
stocks.

$\bullet$ The {\em equal-weighted index} of $\BB$ at time $t$ is
\[\Phi_3(\BB,t)=\sum_{i=1}^b\frac{S_{i,t}}{S_{i,0}}.\]
The index published by the Indicator Digest is calculated by this method,
involving stocks listed on the New York Stock Exchange.

$\bullet$ The {\em price-relative index} of $\BB$ at time $t$ is
\[\Phi_4(\BB,t) =
\left(\Pi_{i=1}^b\frac{S_{i,t}}{S_{i,0}}\right)^{\frac{1}{b}}.\] The Value
Line Index is computed by this formula.

There are numerous reasons why stock investors and money managers
would want to invest in a subset of stocks rather than those of a
whole market \cite{asb93}.  For instance, small investors certainly do
not have sufficient capital to invest in every stock in the market.
Logically, such investors would attempt to choose a small subset of
stocks which hopefully can perform roughly as well as or even
outperform the market as a whole.  They then face difficult trade-offs
between returns and risks.  For these and other reasons of
optimization, we formulate three natural computational problems for the
design of market indices.  Given a market $\MM$ consisting of $m$
stocks, we wish to choose a subset $\MM_k$ of at most $k$ stocks and
calculate an index of $\MM_k$, which is called a {\em $k$-proxy} of
the corresponding index of the whole market $\MM$ (we sometimes refer
to $\MM_k$ as a {\em portfolio}).  Our goal is to choose $\MM_k$ so
that the resulting $k$-proxy tracks or outperforms the corresponding
index of $\MM$.  This paper shows that designing proxies for the above
four indices based on historical data is computationally hard.

We note here that while our problem statements might sound rather
restrictive since error bounds must be met for every time step, we can
use simple padding arguments to extend all of our proofs to more
relaxed problems of the form ``can the error bound be met $x$ percent
of the time?''

\section{Problem Formulations}

\newcommand{\RR}[0]{\overline{R}}

In this section we formally define three basic problems related to
selecting $k$-proxies, or portfolios.

\begin{problem}[tracking an index]\rm\
\label{prob_one}

\begin{itemize}
\item[] {\INPUT}: A market $\MM$ of $m$ stocks, their prices
$S_{i,t}\geq 0$
  for $t=0,\ldots,f$, their numbers $w_i$ of outstanding shares, a real
  $\epsilon_1 > 0$, an integer $k>0$, and some $j \in \{1,2,3,4\}$ to
  indicate the desired type of index.
\item[] {\OUTPUT}: A subset $\MM_k$ of at most $k$ stocks in $\MM$ such
  that 
\begin{equation}
  \left|\frac{\Phi_j(\MM_k,t)}{\Phi_j(\MM_k,0)}
    - \frac{\Phi_j(\MM,t)}{\Phi_j(\MM,0)}\right| \leq
    \epsilon_1{\cdot}\frac{\Phi_j(\MM,t)}{\Phi_j(\MM,0)}
      \mbox{ for all }t=1,\ldots,f.
\label{eq:trackdef}
\end{equation}
\end{itemize}
\end{problem}

\begin{problem}[outperforming an index]\rm\
\label{prob_two}

\begin{itemize}
\item[] {\INPUT}: A market $\MM$ of $m$ stocks, their prices
$S_{i,t}\geq 0$
  for $t=0,\ldots,f$, their numbers $w_i$ of outstanding shares, a real
  $\epsilon_2 \geq 0$, an integer $k>0$, and some $j \in \{1,2,3,4\}$ to
  indicate the desired type of index.
\item[] {\OUTPUT}: A subset $\MM_k$ of at most $k$ stocks in $\MM$ such
  that 
\begin{equation}
   \frac{\Phi_j(\MM_k,t)}{\Phi_j(\MM_k,0)} \geq
     (1+\epsilon_2){\cdot}\frac{\Phi_j(\MM,t)}{\Phi_j(\MM,0)}
        \mbox{ for all } t=1,\ldots,f.
\label{eq:beatdef}
\end{equation}
\end{itemize}
\end{problem}

For the final problem, we need a few extra definitions in order to
analyze the {\em volatility} of a set of stocks.
Let $\BB$ be a set of stocks as defined in \S\ref{sec:intro}.  

$\bullet$ The {\em one-period return} of $\Phi_j$ for $\BB$ at time
$t\geq1$ is
\[R_j(\BB,t)=\ln \frac{\Phi_j(\BB,t)}{\Phi_j(\BB,t-1)}.\]

$\bullet$ The {\em average return} of $\Phi_j$ for $\BB$ up to time
$t\geq1$ is
\[\RR_j(\BB,t)=\frac{\sum_{i=1}^tR_j(\BB,i)}{t}.\]

$\bullet$ The {\em volatility} of $\Phi_j$ for $\BB$ up to time $t\geq2$ is
\[\Delta_j(\BB,t) =
\sqrt{\frac{\sum_{i=1}^t\left(R_j(\BB,i)-\RR_j(\BB,t)\right)^2}{t-1}}.\] 

\begin{problem}[sacrificing return for less volatility]\rm\ 
\label{prob_three}

\begin{itemize}
\item[] {\INPUT}: A market $\MM$ of $m$ stocks, their prices
$S_{i,t}\geq 0$
  for $t=0,\ldots,f$, their numbers $w_i$ of outstanding shares, two reals
  $\alpha, \beta > 0$, an integer $k>0$, and some $j \in \{1,2,3,4\}$ to
  indicate the desired type of index.
\item[] {\OUTPUT}: A subset $\MM_k$ of at most $k$ stocks in $\MM$ such
  that 
\begin{equation}
   \frac{\Phi_j(\MM_k,t)}{\Phi_j(\MM_k,0)} \geq
     \alpha{\cdot}\frac{\Phi_j(\MM,t)}{\Phi_j(\MM,0)}
       \mbox{ for all } t=1,\ldots,f;
\label{eq:perfbounddef}
\end{equation}
\begin{equation}
\Delta_j(\MM_k,s) \leq \beta{\cdot}\Delta_j(\MM,s) 
  \mbox{ for all } s=2,\ldots,f.
\label{eq:volbounddef}
\end{equation}
In this problem, (\ref{eq:perfbounddef}) is called the performance
bound, and (\ref{eq:volbounddef}) is called the volatility bound.
\end{itemize}
\end{problem}

\section{Price-weighted Index}

In this section, we consider taking the value of the market and
portfolio using a price-weighted index, defined in~(\ref{eq:phi1def}).
As given in the problem statements, we use the notation $\Phi_1({\cal
M},t)$ to denote the market average at timestep $t$, and $\Phi_1({\cal
M}_k,t)$ to denote the average of the portfolio at that timestep.

\subsection{Tracking an index}
\label{sec:track}

To solve the problem of tracking the market average, we need to
satisfy (\ref{eq:trackdef}) using function $\Phi_1({\cal B},t)$.  We
will refer to this bound as the ``tracking bound.''  In the following
proofs, we show this by proving an equivalent relation:
\begin{equation}
    1-\epsilon \leq \frac{\Phi_1({\cal M},0)}{\Phi_1({\cal M}_k,0)}
               \cdot\frac{\Phi_1({\cal M}_k,t)}{\Phi_1({\cal M},t)} 
    \leq 1+\epsilon . 
\label{eq:equivbnd}
\end{equation}

\begin{theorem}
\label{thm:trackpricewindex}
Let $\epsilon$ be any error bound satisfying $0<\epsilon<1$ and
specified using $n^{O(1)}$ bits in fixed point notation.  Then the
tracking problem for a price-weighted index with error bound
$\epsilon$ is NP-hard.
\end{theorem}

In the remainder of this section, we prove this theorem by reduction
from the minimum set cover problem.  We will use the notation from the
minimum cover definition given in the classic book on NP-completeness
by Garey and Johnson~\cite{gj:79}: $C$ is a
collection of subsets of a finite set $S$, and $K$ is the desired
cover size.  Specifically, we want a subcollection $C'\subseteq C$
such that $|C'|\leq K$ and every item $x\in S$ is in some subset from
$C'$.

Let $n=|C|$, and consider making an $n\times |S|$ matrix in which each
column corresponds to a fixed item from $S$, and each row corresponds
to a subset $S'\in C$.  The element in row $i$, column $j$ is some
given value $v_1$ if the element in $S$ for that column is in the
subset $S'$, and value $v_0$ if it is not.  Then the minimum cover
problem can be stated as follows: Is there a set of $K$ rows such that
the $K\times |S|$ matrix defined using only those rows has at least
one entry with value $v_1$ in each column?

It makes sense now to consider this $n\times |S|$ matrix as an input
to the portfolio selection problem, where each row corresponds to a
stock and each column corresponds to a timestep, and we are to
choose a portfolio of size $k=K$.  Selecting a
portfolio is then equivalent to selecting the subcollection in the
minimum cover problem.  A subcollection that is missing some item from
$S$ corresponds to a portfolio in which some timestep has all values
equal to $v_0$, and hence the portfolio average at that timestep must
be $v_0$.  Ideally, we would select $v_0$ and $v_1$ in such a way that
the required tracking bound is met if any $v_1$ values are included in
the portfolio, but not if all values are $v_0$.  However, this simple
construction has very unpredictable market averages at each time step,
so we need a slightly more involved construction.

\begin{figure}
\begin{center}
\leavevmode\epsfbox{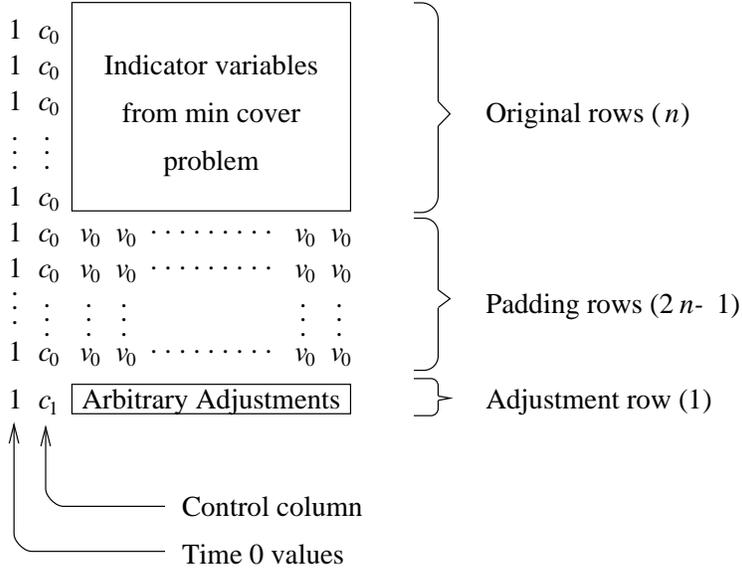}
\end{center}
\caption{Pictorial depiction of reduction for
Theorem~\ref{thm:trackpricewindex}}
\label{fig:reduction}
\end{figure}

We will introduce a new row into our matrix called the ``adjustment
row'', and we will select values to adjust the column averages to
predictable values.  To guarantee that this row is not selected in our
portfolio (so selections are made up entirely of rows from the minimum
cover problem), we introduce a special column called the ``control
column'' --- any selection including our adjustment row will violate
the error bound in that column, and no selection excluding that row
will violate the bound.  In addition, we need to pad the problem out
substantially.  This is accomplished by including rows that contain
value $v_0$ in every non-control column, which is equivalent to
padding the original set cover problem instance with empty subsets
added to $C$.  This clearly has no effect on the set cover problem.
Finally, we insert a column of all ones to give the $S_{i,0}$ values
for the portfolio selection problem.  The final matrix contains $m=3n$
rows, $f=|S|+2$ columns, and is depicted in
Figure~\ref{fig:reduction}.

Note that since $S_{i,0}=1$ for all $i$, $\Phi_1({\cal
M},0)=\Phi_1({\cal M}_k,0)=1$, and so (\ref{eq:equivbnd}) reduces to
just checking that
\[
    1-\epsilon \leq \frac{\Phi_1({\cal M}_k,t)}{\Phi_1({\cal M},t)} 
    \leq 1+\epsilon . 
\]

First we examine properties of the control column, where the values in
that column are defined by
\begin{eqnarray*}
c_0 & = & \left\lceil\frac{1-\epsilon}{\epsilon}\right\rceil , \\
c_1 & = & c_0 + m . 
\end{eqnarray*}

\begin{lemma}
The tracking bound is met for the control column if and only if the
adjustment row is not included in the portfolio.
\end{lemma}

\Proof From the values for $c_0$ and $c_1$, it is clear that the
average value of the control column is $c_0+1$.  Since we will be
examining the error of approximations relative to this average, we
first note that we can bound (due to the ceiling involved in the
definition of $c_0$)
\begin{equation}
\label{eq:avgbound}
\frac{\epsilon}{1+\epsilon} < \frac{1}{c_0+1} \leq \epsilon .
\end{equation}
Any portfolio that does not include the adjustment row has average
value $c_0$, and so we can lower bound the relative error by
\[  \frac{\Phi_1({\cal M}_k,t)}{\Phi_1({\cal M},t)} =
\frac{c_0}{c_0+1} = 1 - \frac{1}{c_0+1} \geq 1-\epsilon . \]
Since the relative error is clearly less than one, it falls into the
acceptable range of values.

On the other hand, if a portfolio {\em does} include the adjustment
row, then the portfolio average is $c_0+m/k$, and so the relative error
is
\[ \frac{\Phi_1({\cal M}_k,t)}{\Phi_1({\cal M},t)} 
= \frac{c_0+m/k}{c_0+1} = 1+\frac{m/k-1}{c_0+1} . \]
Due to our padding of the problem, we know that $k\leq m/3$, and so
$m/k-1\geq 2$.  Using this observation and the bound
from~(\ref{eq:avgbound}) leads to the conclusion that
\[  \frac{\Phi_1({\cal M}_k,t)}{\Phi_1({\cal M},t)} \geq 1+\frac{2}{c_0+1} 
    > 1+\frac{2}{1+\epsilon}\,\epsilon > 1+\epsilon .
\]
In other words, any portfolio that includes the adjustment row will
not meet the required error bound.  Combined with our previous
observation, this completes the proof of the lemma.  \QED

Next we must define the values $v_0$ and $v_1$, and show the
equivalence of our portfolio selection instance with the original set
cover instance.  To do so, define
\begin{eqnarray*}
v_0 & = &
\left\lceil k\frac{1-\epsilon}{\epsilon}\right\rceil , \\
\Delta & = & k+\left\lceil\frac{\epsilon}{1-\epsilon}\right\rceil , \\
v_1 & = & v_0 + \Delta .\\
\end{eqnarray*}
Note that since $\epsilon<1$, all these values are clearly
non-negative integers, as required by the portfolio selection problem.

For column $t$, if there are $M_t$ rows with value $v_1$, then the value
we use in the adjustment row for that column is
\[  A_t = v_0 + (m-M_t)\Delta\ , \]
which is clearly a positive integer, since $M_t<m$.
The sum down the column is
\begin{eqnarray*}
(m-M_t-1) v_0 + M_t v_1 + A_t
   & = & (m-M_t-1) v_0 + M_t (v_0+\Delta) + v_0 + (m-M_t)\Delta\\
   & = & m v_0 + m\Delta ,
\end{eqnarray*}
which means that the column average is $v_0 + \Delta$, or just $v_1$.
Notice the independence from $t$.  We make such an adjustment for
every column in the matrix.

We next demonstrate the equivalence of the produced portfolio
selection instance with the original set cover instance.

\begin{lemma}
The relative error bound is met if and only if the portfolio
contains at least one $v_1$ value in each column (other than the
control column).
\end{lemma}

\Proof First, for the ``only if'' part of the lemma, consider the case
where the relative error bound is met.  Consider any specific column
$t$ of our table, and assume that this column does not contain any
$v_1$ values.  By the last lemma, the adjustment row cannot be
included in our portfolio, so all values must be $v_0$, and so the
portfolio average is exactly $v_0$.  Therefore, we can derive
\begin{equation}
\label{eq:rationov1}
  \frac{\Phi_1(\MM_k,t)}{\Phi_1(\MM,t)} 
     = \frac{v_0}{v_0+\Delta}
     = \frac{1}{1+\frac{\Delta}{v_0}} ,
\end{equation}
and so providing a good lower bound for $\frac{\Delta}{v_0}$ would in
fact upper bound this ratio.  We can do this as follows:
\[
  \frac{\Delta}{v_0} 
    = \frac{k+\left\lceil\frac{\epsilon}{1-\epsilon}\right\rceil}
           {\left\lceil k \frac{1-\epsilon}{\epsilon}\right\rceil}
    > \frac{k+\frac{\epsilon}{1-\epsilon}}
           {k \frac{1-\epsilon}{\epsilon}+1}
    = \frac{\frac{k(1-\epsilon)+\epsilon}{1-\epsilon}}
           {\frac{k(1-\epsilon)+\epsilon}{\epsilon}}
    = \frac{\epsilon}{1-\epsilon} .
\]
Plugging back in to~(\ref{eq:rationov1}), we get
\[
  \frac{\Phi_1(\MM_k,t)}{\Phi_1(\MM,t)} 
     < \frac{1}{1+\frac{\epsilon}{1-\epsilon}}
     = 1-\epsilon .
\]
Thus under our assumption that no $v_1$ values are included, the error
bound is not met.  We conclude that if the error bound is met, then at
least one $v_1$ value must be included in each column.

Next, for the ``if'' part of the theorem, assume that each column in
the selected portfolio contains at least one $v_1$ value and that we
have not selected the adjustment row.  Since the market average is
$v_1$, and the largest possible selected value in the portfolio is
$v_1$, we know that $\Phi_1(\MM_k,t)\leq\Phi_1(\MM,t)$, and so the
upper bound $1+\epsilon$ on the relative error is trivially met for
any $\epsilon\geq 0$.

Since we have selected at least one $v_1$ value, the portfolio average
is at least $v_0+\Delta/k$, and so to lower bound the relative error
notice that
\begin{equation}
\label{eq:boundonev1}
  \frac{\Phi_1(\MM_k,t)}{\Phi_1(\MM,t)} 
    \geq \frac{v_0+\frac{\Delta}{k}}{v_0+\Delta}
    = 1-\frac{k-1}{k}\cdot\frac{1}{\frac{v_0}{\Delta}+1} .
\end{equation}
Now we will derive a lower bound for $\frac{v_0}{\Delta}$ in a similar
way to what we did above, so
\[
   \frac{v_0}{\Delta}
    = \frac{\left\lceil k \frac{1-\epsilon}{\epsilon}\right\rceil}
           {k+\left\lceil\frac{\epsilon}{1-\epsilon}\right\rceil}
    > \frac{k \frac{1-\epsilon}{\epsilon}}
           {k+\frac{\epsilon}{1-\epsilon}+1}
    = \frac{\frac{k(1-\epsilon)}{\epsilon}}
           {\frac{k(1-\epsilon)+1}{1-\epsilon}}
    = \frac{1-\epsilon}{\epsilon}\cdot\frac{k(1-\epsilon)}{k(1-\epsilon)+1} .
\]
Using this bound, with a little manipulation we can derive
\[
   \frac{k-1}{k}\cdot\frac{1}{\frac{v_0}{\Delta}+1}
    < \frac{k-1}{k}\cdot
	\frac{(1-\epsilon)k+1}{(1-\epsilon)k+\epsilon}\cdot\epsilon .
\]
We can bound the middle factor of this bound by $\frac{k+1}{k}$ by
noticing that
\[\def\impliedby{\mbox{\hspace*{0.2in}$\Longleftarrow$\hspace*{0.2in}}}
  \frac{(1-\epsilon)k+1}{(1-\epsilon)k+\epsilon}<\frac{k+1}{k}
  \impliedby
  (1-\epsilon)k^2+k < (1-\epsilon)k^2 + \epsilon k +(1-\epsilon)k +\epsilon
  \impliedby
  0 < \epsilon ,
\]
and so plugging back into~(\ref{eq:boundonev1}) we get
\[
  \frac{\Phi_1(\MM_k,t)}{\Phi_1(\MM,t)} 
    \geq 1-\frac{1}{\frac{v_0}{\Delta}+1}
    > 1-\frac{k-1}{k}\cdot\frac{k+1}{k}\cdot\epsilon
    = 1-\frac{k^2-1}{k^2}\cdot\epsilon > 1-\epsilon .
\]
We conclude that if at least one value in column $t$ of the selected
portfolio is $v_1$, then the relative error bound is met.  Since we
have completed both directions of the ``if and only if'' proof, this
completes the proof of the lemma.
\QED

As a final note, it is fairly easy to show that all values in the
constructed portfolio selection problem have length polynomial in the
length of the original set cover problem and the number of bits used
to specify $\epsilon$.  Therefore, these values form a polynomial time
reduction from the set cover problem to the portfolio selection
problem, which completes the proof of
Theorem~\ref{thm:trackpricewindex}.

\subsection{Sacrificing Return for Less Volatility}
\label{sec:srlv}

Next, we will skip Problem 2 and prove a hardness result for Problem
3:  sacrificing return for less volatility.  In the following section,
we will return to problem 2, and show that the hardness of that
problem (outperforming an index) follows directly from the results of
this section.

As in \S\ref{sec:track}, we will show that Problem 3 is NP-complete
by reducing the minimum cover problem to this one.

\subsubsection{The construction}

\begin{figure}
\begin{center}
\leavevmode\epsfbox{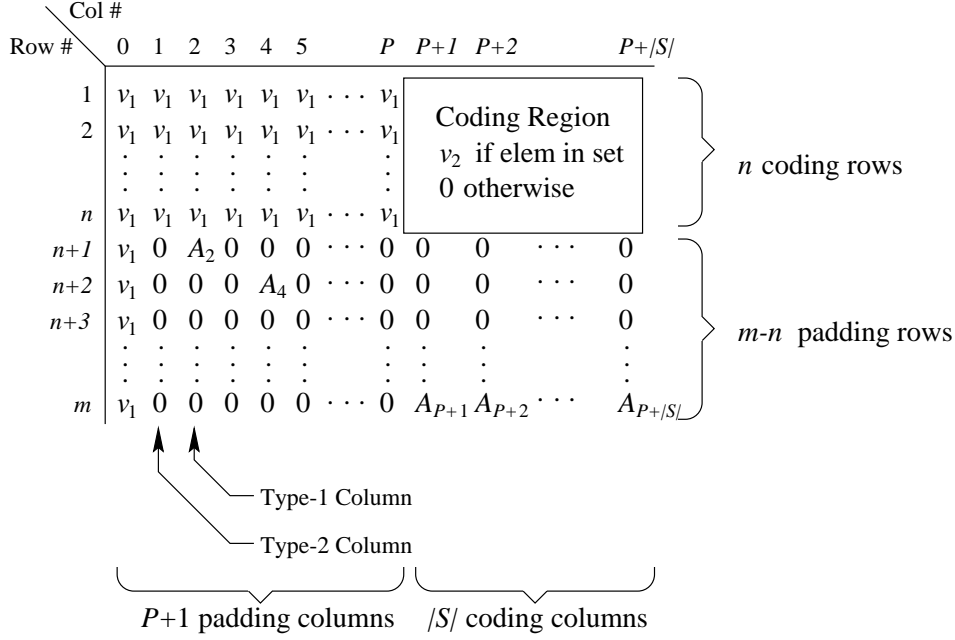}
\end{center}
\caption{Construction for main reduction of Section \ref{sec:srlv}}
\label{fig:srlv}
\end{figure}

The main reduction for this proof involves a problem constructed from
a minimum cover instance, and this construction is illustrated in
Figure~\ref{fig:srlv}.  This constructed problem is an instance of our
portfolio selection problem where the rows represent different
stocks, the columns represent times, and the values in the matrix
represent prices.

In the original minimum cover instance, let $n=|C|$ represent the
number of subsets in the input, let $|S|$ represent the size of
the overall set, and let $K$ be the number of subsets we are allowed
to select.  The data from this problem can be encoded into an $n\times |S|$
matrix $M$, where the values in this matrix are set as follows ($v_2$
is a value that will be defined shortly):
\[  M_{ij} = \left\{ \begin{array}{ll}
      v_2 & \mbox{ if subset $i$ contains element $j$;} \\
      0 & \mbox{ otherwise.}
     \end{array}\right.
\]
We will need a larger matrix in order to complete the reduction, so
we embed matrix $M$ into our larger matrix --- in
Figure~\ref{fig:srlv} the embedded matrix is labeled as the ``Coding
Region''.  This gives a portfolio selection problem with $m$
stocks, $f=P+|S|$ time steps, and portfolio size $k=K$.

We surround matrix $M$ with various ``padding rows'' and ``padding
columns''.  The number of padding rows and padding columns are defined
as follows:
\begin{itemize}
\item There are $P+1$ padding columns, where 
$P=\max\left( 2(k+1), 2|S|\right)$.
\item The total number of rows is defined in terms of the following
constants:
\[ q=\left\lceil\max\left(1+(4/\beta),
\log_k(2/\alpha)\right)\right\rceil , \hbox to 1in{\hss and\hss}
B=\left\lceil \alpha k^q\right\rceil .
\]
The total number of rows is $m=nB$.
\end{itemize}
The definition of $q$ implies some important properties of the
constant $B$ that we note here:
\begin{equation}
B\geq 2 ;
\label{eq:bgeq2}
\end{equation}
\begin{equation}
B\geq k\alpha\geq \alpha .
\label{eq:bgeqalpha}
\end{equation}
Finally, from the first part of~(\ref{eq:bgeqalpha}) we can derive
\begin{equation}
\left\lfloor\frac{B}{\alpha}\right\rfloor > \frac{B}{\alpha}\,
\frac{k-1}{k} .
\label{eq:bdivalphabound}
\end{equation}

All of the first $n$ rows in the padding columns are filled with value
$v_1$, and value $v_2$ is used in the coding region as previously
described.  These values are defined in terms of the constant $B$ as
follows:
\begin{itemize}
\item $v_1=B-1$
\item $v_2=k(B-1)$
\end{itemize}

Each column may have an ``adjustment value'', denoted by $A_t$ for
column $t$.  Odd numbered columns in the padding region (type-2
columns) do not have an adjustment value, but even numbered columns
other than column 0 (type-1 columns) do, and these values are
positioned at successively lower rows; therefore, if column $t$ is a
type-1 column, then $A_t$ is placed in row $n+\frac{t}{2}$.  If we run
out of rows before completing this placement, simply put all remaining
adjustment values on the last row.  Notice that since $P\geq 2(k+1)$
there are at least $k+1$ type-1 padding columns, and since the number
of padding rows is $(m-n)=(nB-n)\geq n\geq k+1$
(using~(\ref{eq:bgeq2})), there must be at least $k+1$ distinct rows
that contain adjustment values.  Columns that cross the coding region
(called ``coding columns'') also have adjustment values, which are all
placed on the last row of the matrix (see Figure~\ref{fig:srlv}).  The
adjustment values to be used are defined below, where $z_t$ is the
number of zeros in the coding region of column $t$:
\[ A_t = \left\{ \begin{array}{ll}
    (m-n)\left(\left\lfloor\frac{B}{\alpha}\right\rfloor-1\right) & 
	\mbox{ if $0<t\leq P$ and $t$ is even;} \\
\ \\
    (m-n)\left(\left\lfloor\frac{B}{\alpha}\right\rfloor-k\right)+z_t\cdot v_2 & \mbox{ if $t>P$.}
    \end{array}\right. 
\]
Note that the adjustment values in the padding columns are all the
same, but the adjustments in the coding region depend on the data in
the coding region.  Furthermore, (\ref{eq:bgeqalpha}) guarantees
that these adjustment values are all non-negative.

Before analyzing the return and volatility of the constructed
portfolio selection problem, we state the following lemma regarding
the size of the constructed problem, showing that we have a polynomial
reduction --- the proof of this lemma is straight-forward given the
above definitions, and is omitted.

\begin{lemma}
\label{lem:srlv_size}
If $\alpha$ and $\beta$ are expressed using $n^{O(1)}$ bits in
fixed-point binary notation, and $0<\alpha\leq n^{O(1)}$ and
$\beta=\Omega\left(\frac{\log k}{\log n}\right)$, then the size of the
constructed problem (including the size of the values in the matrix)
is polynomial in the size of the original minimum cover problem.
\end{lemma}

\subsubsection{Guarantees on Return}

\begin{lemma}
\label{lem:srlv_return}
The performance bound is met for all columns if and only if the
selected portfolio contains exactly $k$ items from the coding rows and
each coding column has at least one $v_2$ value from among the
selected rows.
\end{lemma}

\Proof We will first prove that if the selected portfolio contains
exactly $k$ items from the coding rows and each coding column has at
least one $v_2$ value from the selected rows, then the performance
bound is met.  First consider a padding column $t$ --- since the
$k$ selected rows are all coding rows, all selected values for any
padding column have value $v_1$, and so the portfolio average for that
column is $\Phi_1({\cal M}_k,t)=v_1$.  On the other hand, the market
average is different for the two types of columns.  If column $t$ is a
type-1 padding column, then the sum of all the values in the column is
\begin{eqnarray*}
n v_1 + A_t & = & n (B-1) + 
    (m-n)\left(\left\lfloor\frac{B}{\alpha}\right\rfloor -1\right)
   = n (B-1) +
    (nB-n)\left(\left\lfloor\frac{B}{\alpha}\right\rfloor -1\right) \\
   & = & n (B-1) +
    (B-1)\left(n\left\lfloor\frac{B}{\alpha}\right\rfloor -n\right)
   = (B-1)n\left\lfloor\frac{B}{\alpha}\right\rfloor .
\end{eqnarray*}
Therefore, the market average for column $t$ satisfies
\begin{eqnarray}
\Phi_1({\cal M},t) & = &
   \frac{(B-1)n}{nB}\left\lfloor\frac{B}{\alpha}\right\rfloor 
   = \frac{B-1}{B}\,\left\lfloor\frac{B}{\alpha}\right\rfloor 
\label{eq:type1ma}\\
   & \leq & \frac{B-1}{B}\,\frac{B}{\alpha} = \frac{B-1}{\alpha}
   = \frac{v_1}{\alpha} .
\nonumber
\end{eqnarray}
Furthermore, any type-2 padding column has no adjustment value, which
makes the market average smaller than a type-1 column.  Therefore, for
either type of padding column the bound $\Phi_1({\cal M},t)\leq
\frac{v_1}{\alpha}$ is valid, and so it immediately follows that for
any padding column $t$, since $\Phi_1({\cal M},0)=\Phi_1({\cal
M}_k,0)=v_1$,
\[ \frac{\Phi_1({\cal M}_k,t)}{\Phi_1({\cal M}_k,0)} 
   \geq \alpha\cdot\frac{\Phi_1({\cal M},t)}{\Phi_1({\cal M},0)} .
\]
Therefore, the performance bound is met for all padding columns.

Now consider a coding column $t$, and recall that we are assuming that
at least one $v_2$ value from column $t$ is included in the portfolio.
This means that the portfolio average is $\Phi_1({\cal M}_k,t)\geq
v_2/k = v_1$.  For the market average, we compute the sum over all
values in the column, as we did before, and in this case we get
\begin{eqnarray*}
(n-z_t)v_2 + A_t & = &
n v_2 - z_t v_2 + (m-n)\left(\left\lfloor\frac{B}{\alpha}\right\rfloor
  -k\right) + z_t v_2 \\
   & = & n k(B-1) +
    (nB-n)\left(\left\lfloor\frac{B}{\alpha}\right\rfloor -k\right) \\
   & = & n k(B-1) +
    (B-1)\left(n\left\lfloor\frac{B}{\alpha}\right\rfloor - nk\right)
   = (B-1)n\left\lfloor\frac{B}{\alpha}\right\rfloor .
\end{eqnarray*}
Similar to the calculation for the padding columns, this gives us
\begin{equation}
\label{eq:codingma}
\Phi_1({\cal M},t) =
   \frac{B-1}{B}\,\left\lfloor\frac{B}{\alpha}\right\rfloor 
	\leq \frac{B-1}{\alpha}
   = \frac{v_1}{\alpha} \hbox to 1in{\hss$\Longrightarrow$\hss}
\frac{\Phi_1({\cal M}_k,t)}{\Phi_1({\cal M}_k,0)} \geq 
\alpha\cdot\frac{\Phi_1({\cal M},t)}{\Phi_1({\cal M},0)} ,
\end{equation}
and so the performance bound is met for the coding columns as well.
Therefore we have completed this direction of the proof.

For the other direction, we need to show that any portfolio that meets
the performance bound must be made up of exactly $k$ items from the
coding rows and each coding column has at least one $v_2$ value from
the selected rows.  We first show that any portfolio that meets the
performance bound may only use coding rows.  By our placement of
adjustment values, we noticed before that there are at least $k+1$
distinct padding rows that contain adjustment values.
Therefore, there must be at least one type-1 padding column, say
column $t$, that does not have its adjustment value $A_t$ selected as part
of the portfolio.  Now if all $k$ selections are {\em not} from the coding
rows, then we can bound the portfolio average for column $t$ by
\[ \Phi_1({\cal M}_k,t) \leq \frac{(k-1)v_1}{k} . \]
Since this is a type-1 column, (\ref{eq:type1ma}) gives the market
average, and we can further use~(\ref{eq:bdivalphabound}) to conclude that
\[ \frac{\Phi_1({\cal M}_k,t)}{\Phi_1({\cal M}_k,0)}\,
   \frac{\Phi_1({\cal M},0)}{\Phi_1({\cal M},t)}
    \leq \frac{\frac{(k-1)v_1}{k}}{v_1}\,
     \frac{v_1}{\frac{(B-1)}{B}\left\lfloor\frac{B}{\alpha}\right\rfloor}
    < \frac{(k-1)(B-1)}{k}\,
     \frac{1}{\frac{(B-1)}{B}\frac{B}{\alpha}\,\frac{k-1}{k}}
    = \alpha ,
\]
and so the performance bound would not be met.  Therefore, all $k$ row
selections must come from the coding rows.

Since we have established that all $k$ selections must come from the
coding rows, we will next show that every column in the coding region
must have at least one $v_2$ value among the selected rows.  This is,
in fact, very easy to see --- if no $v_2$ values are selected in a
particular column, then the portfolio average is zero, which cannot
meet the performance bound for that column.  Therefore, all coding
columns must be contain at least one $v_2$ value, which completes this
direction of the proof, and also completes the entire proof.  \QED

\subsubsection{Guarantees on Volatility}

\begin{lemma}
\label{lem:srlv_volatility}
If the performance bound is met for our constructed portfolio
selection problem, then the volatility bound is met as well.
\end{lemma}

\Proof
Assume we have a solution that meets the performance bounds.
Then by Lemma~\ref{lem:srlv_return} we know that all $k$
selected rows are coding rows and that each coding column contains at
least one $v_2$ value.  From this information, we can bound the
volatility of both the market and the portfolio.

The first observation is that the portfolio average is exactly $v_1$
for every padding column, including column 0, and this constant
average means that the portfolio volatility is exactly zero for all of
the padding columns (so $\Delta({\cal M}_k,t)=0$ for all $t\leq P$).
Since the portfolio volatility is zero, the volatility bound is
trivially met whenever $t\leq P$.

For $t>P$ we bound the market volatilities first.  We have already
computed the market averages for the type-1 columns
(in~(\ref{eq:type1ma})) and for coding columns
(in~(\ref{eq:codingma})), but we need to compute the market average
for type-2 columns.  Since there are exactly $n$ values of $v_1$ in a
type-2 column, and there are $m=nB$ total columns, the market average
of a type-2 column is simply
$\frac{nv_1}{m}=\frac{n(B-1)}{nB}=\frac{B-1}{B}$.  We summarize all
market averages below:
\[  \Phi_1({\cal M},t) = \left\{ \begin{array}{ll}
	B-1 & \mbox{ if $t=0$;}\\
\ \\
	\frac{B-1}{B} & \mbox{ if $t\leq P$ and $t$ is odd;} \\
\ \\
	\frac{B-1}{B}\left\lfloor\frac{B}{\alpha}\right\rfloor &
		\mbox{ otherwise.}
   \end{array}\right.
\]
These values can then be used to compute the one-period returns for
the market:
\[  R_1({\cal M},i) = \left\{\begin{array}{ll}
	-\ln B & \mbox{ if $i=1$;}\\
\ \\
	-\ln\left\lfloor\frac{B}{\alpha}\right\rfloor & 
		\mbox{ if $1<i\leq P$ and $i$ is odd;} \\
\ \\
	\ln\left\lfloor\frac{B}{\alpha}\right\rfloor & 
		\mbox{ if $i\leq P$ and $i$ is even;} \\
\ \\
	0 & \mbox{ if $i>P$.}
	\end{array}\right.
\]

Recall that we are only interested in volatilities for times $t>P$,
and from the above we can derive for $t>P$
\[ \RR_1({\cal M},t) = 
   \frac{1}{t}\ln\frac{\Phi_1({\cal M},t)}{\Phi_1({\cal M},0)} =
   \frac{1}{t}\ln\left(\frac{1}{B}\,\left\lfloor\frac{B}{\alpha}\right\rfloor\right) .
\]
This market average return can be either positive or negative,
depending on the value of $\alpha$, so we
consider these two situations separately.  First, if $\alpha\geq 1$,
then $B\geq\left\lfloor\frac{B}{\alpha}\right\rfloor$, and so
$\RR_1({\cal M},t)\leq 0$, which implies that when $i$ is even we have
\[  
  R_1({\cal M},i)-\RR_1({\cal M},t) \geq R_1({\cal M},i)
    = \ln\left\lfloor\frac{B}{\alpha}\right\rfloor
  \hbox to 0.8in{\hss$\Longrightarrow$\hss}
  \left(R_1({\cal M},i)-\RR_1({\cal M},t)\right)^2 \geq 
     \left(\ln\left\lfloor\frac{B}{\alpha}\right\rfloor\right)^2 .
\]
On the other hand, if $\alpha<1$, 
then $B<\left\lfloor\frac{B}{\alpha}\right\rfloor$, and so
$\RR_1({\cal M},t)> 0$, which implies that when $i$ is odd and greater
than 1 we have
\[  
  R_1({\cal M},i)-\RR_1({\cal M},t) \leq R_1({\cal M},i)
    = -\ln\left\lfloor\frac{B}{\alpha}\right\rfloor
  \hbox to 0.8in{\hss$\Longrightarrow$\hss}
  \left(R_1({\cal M},i)-\RR_1({\cal M},t)\right)^2 \geq 
     \left(\ln\left\lfloor\frac{B}{\alpha}\right\rfloor\right)^2 .
\]
Notice that in both cases, we have the same bound, and we can
guarantee that this bound holds for at least $\frac{P}{2}-1$ columns.
Using this fact, we can bound the market volatilities for $t>P$ as
follows:
\[
\Delta_1({\cal M},t) = 
\sqrt{\frac{\sum_{i=1}^t \left(R_1({\cal M},i)-\RR_1({\cal M},t)\right)^2}
           {t-1}} 
\geq \sqrt{\frac{\left(\frac{P}{2}-1\right)\left(\ln\left\lfloor\frac{B}{\alpha}\right\rfloor\right)^2}
             {t-1}}
= \sqrt{\frac{P-2}{2(t-1)}}\ln\left\lfloor\frac{B}{\alpha}\right\rfloor .
\]
Since $t\leq P+|S|$, $P\geq 2|S|$, and $P\geq 6$, we can bound
$\frac{P-2}{2(t-1)}\geq \frac{1}{4}$, and then
use (\ref{eq:bdivalphabound}) to derive
\begin{eqnarray}
\Delta_1({\cal M},t) 
& = & \sqrt{\frac{1}{4}}\ln\left\lfloor\frac{B}{\alpha}\right\rfloor
\geq \frac{1}{2}\ln\left\lfloor\frac{B}{\alpha}\right\rfloor
> \frac{1}{2}\ln\left(\frac{B}{\alpha}\,\frac{k-1}{k}\right)
\geq \frac{1}{2}\ln\left(\frac{\alpha
k^q}{\alpha}\,\frac{k-1}{k}\right)
\nonumber\\
& = & \frac{1}{2}\ln\left(k^q\,\frac{k-1}{k}\right)
\geq \frac{1}{2}\ln\left(k^{(4/\beta)+1}\,\frac{k-1}{k}\right)
= \frac{1}{2}\ln\left(k^{(4/\beta)}(k-1)\right) \nonumber\\
& \geq & \frac{1}{2}\ln k^{(4/\beta)}
= \frac{1}{2}\,\frac{4}{\beta}\,\ln k 
> \frac{2}{\beta}\,\ln k .
\label{eq:marketvol}
\end{eqnarray}

Next, we will find an upper bound for the portfolio volatility.  As
mentioned before, the portfolio averages for $t\leq P$ are constant
values $v_1$.  For $t>P$, the portfolio averages are data dependent,
but we can certainly bound them by the closed interval
\[  \Phi_1({\cal M}_k,t)\in [ \frac{v_2}{k},v_2 ] = [ B-1, k(B-1) ] . \]
Using this bound, we can bound the one-period portfolio returns by
\[
  \ln\frac{\Phi_1({\cal M}_k,t)}{\Phi_1({\cal M}_k,t-1)}
   \in [ \ln\frac{B-1}{k(B-1)},\ln\frac{k(B-1)}{B-1} ]
   = [ -\ln k, \ln k ] ,
\]
and we can also bound the portfolio's average return by
\[
  \frac{1}{t} \ln\frac{\Phi_1({\cal M}_k,t)}{\Phi_1({\cal M}_k,0)}
  \in [ \frac{1}{t} \ln\frac{B-1}{B-1}, \frac{1}{t} \ln\frac{k(B-1)}{B-1} ]
  = [ 0,\frac{1}{t}\ln k] .
\]
Given these bounds, the largest possible value for
$(R_1({\cal M}_k,i)-\RR_1({\cal M}_k,t))^2$ is 
$\left(\frac{t+1}{t}\ln k\right)^2$, and so
\[
\Delta_1({\cal M}_k,t) =
\sqrt{\frac{\sum_{i=1}^t 
   \left(R_1({\cal M}_k,i)-\RR_1({\cal M}_k,t)\right)^2}{t-1}}
\leq \sqrt{\frac{ t\left(\frac{t+1}{t}\right)^2}{t-1}}\,\ln k
 = \sqrt{\frac{(t+1)^2}{t(t-1)}}\,\ln k .
\]
Finally, since $t\geq P+1\geq 2t+1\geq 3$, we can bound
\begin{equation}
\Delta_1({\cal M}_k,t) \leq 2\ln k .
\label{eq:portvol}
\end{equation}

Combining~(\ref{eq:marketvol}) and~(\ref{eq:portvol}) we get
\[ \frac{\Delta_1({\cal M}_k,t)}{\Delta_1({\cal M},t)}
   < \frac{2\ln k}{\frac{2}{\beta}\,\ln k} = \beta ,
\]
and so the volatility bounds are met.
\QED

\subsubsection{The main result}

\begin{theorem}
\label{thm:srlv_main}
Let $\alpha$ and $\beta$ be values expressed using $n^{O(1)}$ bits in
fixed-point binary notation, and satisfying $0<\alpha\leq n^{O(1)}$
and $\beta=\Omega\left(\frac{\log k}{\log n}\right)$.  Then the
problem of sacrificing return for less volatility using the
price-weighted index is NP-complete.
\end{theorem}

\Proof Follows immediately from Lemmas~\ref{lem:srlv_size},
\ref{lem:srlv_return}, and \ref{lem:srlv_volatility}. \QED

\subsection{Outperforming an index}

Given the results of the previous section, showing that the problem of
outperforming an index is NP-complete is trivial.  In particular, we
use the exact same construction as in Section~\ref{sec:srlv} (for
concreteness in the construction, use $\beta=4$), and then our result
follows from direct application of Lemmas~\ref{lem:srlv_size}
and~\ref{lem:srlv_return}.

\begin{theorem}
\label{thm:beatpricewindex}
Let $\epsilon$ be any value satisfying $0<\epsilon<n^c$ for some
constant $c$.  Then the problem of outperforming the market average
using the price-weighted index with bound $\epsilon$ is NP-hard.
\end{theorem}

We note here that the construction of Section~\ref{sec:srlv} gives us
a slightly stronger result:  We can actually let $\epsilon$ be as
small as $-1+2^{-n^{O(1)}}$.  However, the disadvantage of using this
reduction is that it is in fact more complicated than necessary for
this problem --- a direct, and simpler, reduction for the problem of
outperforming an index is given in the appendix.

\section{Other Indices}

For the value-weighted and equal-weighted indices, we will, in fact,
use the exact same constructions as in the previous section --- the
prices in the constructed problem have been selected carefully so that
they work using related indices, such as the value-weighted and
equal-weighted indices.  The results will follow fairly easily from
the following lemma.

\begin{lemma}
\label{lem:otherindex}
Let $\Phi_j({\cal B},t)$ be an index function where
$S_{i,0}=c$ for some constant c implies that
\begin{equation}
\label{eq:indexcond}
 \frac{\Phi_j({\cal B},t)}{\Phi_j({\cal B},0)} 
  = d\,\cdot\,\Phi_1({\cal B},t)
\end{equation}
for all sets of stocks ${\cal B}\subseteq{\cal M}$, where $d$ is a
constant that does not depend on ${\cal B}$ or $t$.  Then all of the
previous NP-completeness results hold for index $\Phi_j({\cal B},t)$.
\end{lemma}

\Proof Note that in all the problem statements, whenever an index
value is used, it is {\em always} used in a ratio with the same index
function, either at a different time step or for a different set of
stocks.  This will allow us to cancel out common factors, and the
resulting problem will be in terms of the price-weighted index
($\Phi_1({\cal B},t)$).  For example, in considering the tracking
problem, we need to have a subset ${\cal M}_k$ of $k$ stocks such that
for all $t=1,\ldots,f$,
\[
  \left|\frac{\Phi_j(\MM_k,t)}{\Phi_j(\MM_k,0)}
    - \frac{\Phi_j(\MM,t)}{\Phi_j(\MM,0)}\right| \leq
    \epsilon_1{\cdot}\frac{\Phi_j(\MM,t)}{\Phi_j(\MM,0)} .
\]
Due to the condition of equation~(\ref{eq:indexcond}), this bound is
met if and only if
\[
  \left|\frac{d\cdot\Phi_j(\MM_k,0)\cdot\Phi_1(\MM_k,t)}
	{d\cdot\Phi_j(\MM_k,0)\cdot\Phi_1(\MM_k,0)}
    - \frac{d\cdot\Phi_j(\MM,0)\cdot\Phi_1(\MM,t)}
	{d\cdot\Phi_j(\MM,0)\cdot\Phi_1(\MM,0)}\right| \leq
    \epsilon_1{\cdot}\frac{d\cdot\Phi_j(\MM,0)\cdot\Phi_1(\MM,t)}
	{d\cdot\Phi_j(\MM,0)\cdot\Phi_1(\MM,0)} ,
\]
and cancelling common terms we see that this is met if and only if
\[
  \left|\frac{\Phi_1(\MM_k,t)}{\Phi_1(\MM_k,0)}
    - \frac{\Phi_1(\MM,t)}{\Phi_1(\MM,0)}\right| \leq
    \epsilon_1{\cdot}\frac{\Phi_1(\MM,t)}{\Phi_1(\MM,0)} .
\]
Therefore, the tracking problem using the $\Phi_j$ index function is
entirely equivalent to the problem using the $\Phi_1$ index function.

Exactly the same derivation can be performed on the Problem 2
condition~(\ref{eq:beatdef}), on the definition of $R_j({\cal B},t)$,
and on the Problem 3 performance bound~(\ref{eq:perfbounddef}).
Therefore, all of these problems are equivalent to using the
price-weighted index, and our previous reductions apply.
\QED

\subsection{The Value-Weighted Index}

We first apply Lemma~\ref{lem:otherindex} to the value-weighted index.
For the value-weighted index, we must indicate the weights (the
$w_i$'s) in the constructed portfolio selection problem as well as the
prices.  In all of our constructions, we will pick $w_i=1$ for all
$i$.

If $S_{i,0}=c$ for some constant $c$, then
for any valid time $t$ and any set of stocks ${\cal B}$,
using $w_i=1$ gives
\[
 \Phi_2({\cal B},t) = 
    \frac{\sum_{i=1}^b w_i\cdot S_{i,t}}{\sum_{i=1}^b w_i\cdot S_{i,0}}
  = \frac{\sum_{i=1}^b S_{i,t}}{\sum_{i=1}^b c}
  = \frac{\sum_{i=1}^b S_{i,t}}{b\,c} = \frac{1}{c}\Phi_1({\cal B},t) .
\]
Furthermore, regardless of ${\cal B}$ we have $\Phi_2({\cal B},0)=1$,
and so Lemma~\ref{lem:otherindex} holds with constant $d=\frac{1}{c}$.
The following three theorems are a direct consequence of this Lemma.

\begin{theorem}
Let $\epsilon$ be any error bound satisfying $0<\epsilon<1$ and
specified using $n^{O(1)}$ bits in fixed point notation.  Then the
tracking problem for a value-weighted index with error bound
$\epsilon$ is NP-hard.
\end{theorem}

\begin{theorem}
Let $\epsilon$ be any value satisfying $0<\epsilon<n^c$ for some
constant $c$.  Then the problem of outperforming the market average
using the value-weighted index with bound $\epsilon$ is NP-hard.
\end{theorem}

\begin{theorem}
Let $\alpha$ and $\beta$ be values expressed using $n^{O(1)}$ bits in
fixed-point binary notation, and satisfying $0<\alpha\leq n^{O(1)}$
and $\beta=\Omega\left(\frac{\log k}{\log n}\right)$.  Then the
problem of sacrificing return for less volatility using the
value-weighted index is NP-complete.
\end{theorem}

\subsection{The Equal-Weighted Index}

If $S_{i,0}=c$ for all $i$, then
\[  
  \Phi_3({\cal B},t)
  = \sum_{i=1}^b \frac{S_{i,t}}{S_{i,0}}
  = \sum_{i=1}^b \frac{S_{i,t}}{c} = \frac{1}{c}\sum_{i=1}^b S_{i,t} 
  = \frac{b}{c}\,\Phi_1({\cal B},t) .
\]
It's easy to see that $\Phi_3({\cal B},0)=b$, so
\[
  \frac{\Phi_3({\cal B},t)}{\Phi_3({\cal B},0)}
  = \frac{1}{c}\,\Phi_1({\cal B},t) ,
\]
and so Lemma~\ref{lem:otherindex} applies with constant
$d=\frac{1}{c}$.
The following three theorems are direct consequences of that Lemma.

\begin{theorem}
Let $\epsilon$ be any error bound satisfying $0<\epsilon<1$ and
specified using $n^{O(1)}$ bits in fixed point notation.  Then the
tracking problem for a equal-weighted index with error bound
$\epsilon$ is NP-hard.
\end{theorem}

\begin{theorem}
Let $\epsilon$ be any value satisfying $0<\epsilon<n^c$ for some
constant $c$.  Then the problem of outperforming the market average
using the equal-weighted index with bound $\epsilon$ is NP-hard.
\end{theorem}

\begin{theorem}
Let $\alpha$ and $\beta$ be values expressed using $n^{O(1)}$ bits in
fixed-point binary notation, and satisfying $0<\alpha\leq n^{O(1)}$
and $\beta=\Omega\left(\frac{\log k}{\log n}\right)$.  Then the
problem of sacrificing return for less volatility using the
equal-weighted index is NP-complete.
\end{theorem}

\subsection{The Price-Relative Index}

The price-relative index is a geometric mean of the values in a
set of stocks, whereas our first index (the price-weighted index) is
the arithmetic mean.  In this section we will show that, at least for
the first two problems, we can transform the reductions for the
price-weighted index into reductions for the price-relative index, and
thus obtain NP-hardness results for the price-relative index.  For the
second problem (outperforming an index), we use the simpler reduction
given in the appendix.  We will use the notation $(S,\epsilon,\Phi_j)$
to denote an instance of a portfolio selection problem with prices
$S_{i,t}$, error bound $\epsilon$, and index function $\Phi_j$.

The first step in transforming the reductions for the price-relative
index is to change them so that every column, including the control
column, has the same market average.  If $c_1,c_2,\ldots,c_n$ are the
column sums of columns 1 through $n$, then let
$c=LCM(c_1,\ldots,c_n)$ be the least common multiple of these sums.
We create a new set of prices by setting
$S_{i,t}'=\frac{c}{c_i}S_{i,t}$ at all times $t\geq 1$.  Now the sum
down column $i$ is
\[  \sum_t S_{i,t}' = \sum_t \frac{c}{c_i}S_{i,t}
    = \frac{c}{c_i} \sum_t S_{i,t} = \frac{c}{c_i}\cdot c_i = c ,
\]
which is independent of the actual column, so all columns will now
have the same average value (so $\Phi_1(\MM,t_1)=\Phi_1(\MM,t_2)$ for
all times $t_1$ and $t_2$).  And finally, since the first two problems
treat columns independently and the bounds are relative error bounds,
if all values in a particular column are multiplied by a particular
value, this ``scaling up'' does not change whether or not the error
bound is met.  Therefore, for problem 1 or problem 2, the instance
$(S_{i,t},\epsilon,\Phi_1)$ satisfies the bound if and only if the
instance $(S_{i,t}',\epsilon,\Phi_1)$ satisfies the bound.

The next step in transforming the reductions is to change all the
$S_{i,t}'$ values into new values $S_{i,t}''=2^{S_{i,t}'}$ for $t\geq 1$,
while keeping $S_{i,0}''=1$ for all $i$.  The result of this is
that for any set of $b$ stocks ${\cal B}$, and any $t\geq 1$,
\[ 
  \Phi_4({\cal B}'',t) 
    = \left(\prod_{i=1}^b \frac{S_{i,t}''}{S_{i,0}''}\right)^{1/b}
    = \left(\prod_{i=1}^b 2^{S_{i,t}'}\right)^{1/b}
    = 2^{(1/b)\sum_{i=1}^b S_{i,t}'}
    = 2^{\Phi_1({\cal B}',t)} .
\]
We will also need to transform the $\epsilon$ values, but this is done
differently for the two problems, and so is handled separately below.

\begin{theorem}
Let $\epsilon$ be any error bound satisfying $0<\epsilon<1$ and
specified using $O(\log n)$ bits in fixed point notation.  Then the
tracking problem for a price-relative index with error bound
$\epsilon$ is NP-hard.
\end{theorem}

\Proof Let $\epsilon'=\frac{\lg\frac{1}{1-\epsilon}}{c/m}$, where $m$
is the number of stocks in the entire market (or the number of rows in
our table), and $c$ is the common column sum as described above in the
transformation from $S$ to $S'$.  Now we show that
$(S'',\epsilon,\Phi_4)$ satisfies the tracking lower bound if and only
if $(S,\epsilon',\Phi_1)$ does:
\[
{\renewcommand{\arraystretch}{1.8}
\begin{array}{lrcl}
& \Phi_1(\MM_k,t) & \geq & (1-\epsilon')\Phi_1(\MM,t)\\
\Longleftrightarrow &
\Phi_1(\MM_k,t) & \geq & (1-\frac{\lg\frac{1}{1-\epsilon}}{c/m})\Phi_1(\MM,t)\\
\Longleftrightarrow &
\Phi_1(\MM_k,t) & \geq & 
   (1-\frac{\lg\frac{1}{1-\epsilon}}{\Phi_1(\MM,t)}\Phi_1(\MM,t)\\
\Longleftrightarrow &
\Phi_1(\MM_k,t) & \geq & \Phi_1(\MM,t) - \lg\frac{1}{1-\epsilon}\\
\Longleftrightarrow &
\Phi_1(\MM_k,t) & \geq & \Phi_1(\MM,t) + \lg(1-\epsilon)\\
\Longleftrightarrow &
2^{\Phi_1(\MM_k,t)} & \geq & 2^{\Phi_1(\MM,t) + \lg(1-\epsilon)}\\
\Longleftrightarrow &
2^{\Phi_1(\MM_k,t)} & \geq & (1-\epsilon)2^{\Phi_1(\MM,t)}\\
\Longleftrightarrow &
\Phi_4(\MM_k,t) & \geq & (1-\epsilon)\Phi_4(\MM,t)
\end{array}}
\]
Furthermore, since the $S_{i,t}$ values come from the reduction 
for Theorem~\ref{thm:trackpricewindex}, the tracking upper bound is
trivially met for $(S'',\epsilon,\Phi_4)$ just like it is trivially
met for $(S,\epsilon',\Phi_1)$ (all acceptable portfolio averages are
in fact less than the market average).

Therefore, $(S'',\epsilon,\Phi_4)$ satisfies the tracking bound (both
upper and lower) if and only if $(S,\epsilon',\Phi_1)$ does, and so we
can use $(S'',\epsilon,\Phi_4)$ in the reduction for the tracking
problem in place of $(S,\epsilon',\Phi_1)$, and the validity of the
reduction for $(S'',\epsilon,\Phi_4)$ follows directly from the
results of Theorem~\ref{thm:trackpricewindex}.  Examining the number
of bits required for the various values in the reduction, we get the
NP-completeness result stated in the theorem.
\QED

\begin{theorem}
Let $\epsilon$ be any value satisfying $0<\epsilon<n^c$ for some
constant $c$.  Then the problem of outperforming the market average
using the price-relative index with bound $\epsilon$ is NP-hard.
\end{theorem}

\Proof Similar to the derivation in the previous theorem, except we
use $\epsilon'=\frac{\lg(1+\epsilon)}{c/m}$.  \QED

Finally, we end this section by noting that our final problem,
sacrificing return for less volatility, does not have independent
column values as problems 1 and 2 did, and so the above transformation
idea does not work.  We leave the complexity of the combination of
price-relative index and problem 3 as an open problem.

\bibliographystyle{siam}
\bibliography{portfolio}

\newpage
\appendix

\section{Direct construction for outperforming an index}

We now turn our attention to the problem of finding a portfolio that
outperforms the market average at every time step.  In particular, we
are looking for a portfolio ${\cal M}_k$ of size $k$ which satisfies
(\ref{eq:beatdef}).  As we did in the first construction (for tracking
an index), we rewrite this condition as follows:
\begin{equation}
\frac{\Phi_1({\cal M},0)}{\Phi_1({\cal M}_k,0)}\cdot
\frac{\Phi_1({\cal M}_k,t)}{\Phi_1({\cal M},t)} \geq 1+\epsilon .
\label{eq:beatbnd}
\end{equation}

\begin{theorem}
\label{thm:directbeatpricewindex}
Let $\epsilon$ be any value satisfying $0<\epsilon<n^c$ for some
constant $c$.  Then the problem of portfolio selection for
outperforming the market average with bound $\epsilon$ is NP-hard.
\end{theorem}

\begin{figure}
\begin{center}
\leavevmode\epsfbox{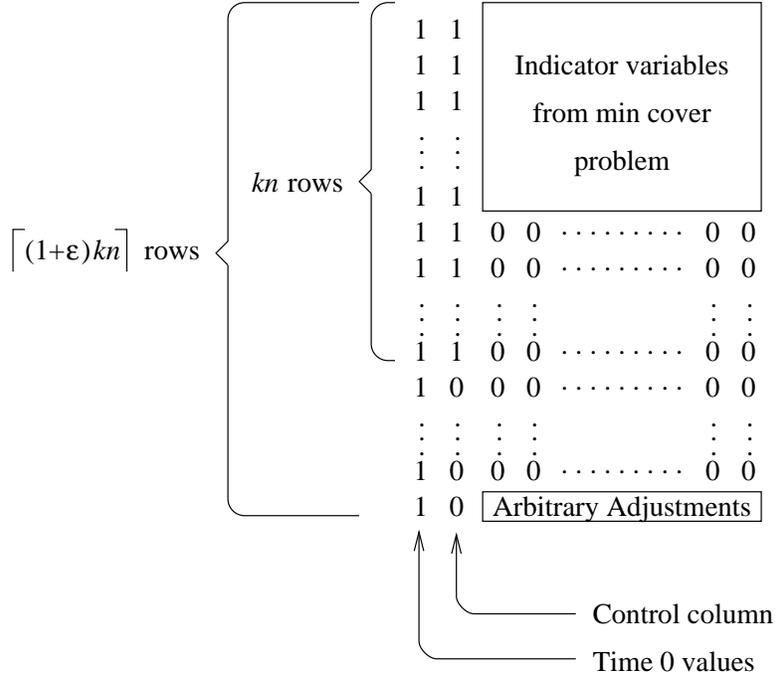}
\end{center}
\caption{Pictorial depiction of reduction for
Theorem~\ref{thm:directbeatpricewindex}}
\label{fig:reduction2}
\end{figure}

\Proof
The reduction used in this proof is shown pictorially in
Figure~\ref{fig:reduction2}.  The indicator variables in this case are
simple zero and one values (set to one if and only if the element
represented by that row is in the subset represented by that column).
The adjustment row contains values so that each column except the
control column has sum $n$.  This is clearly possible for each column,
using only integer values between 0 and $n$.  We also again use an
initial column of all ones, which reduces condition (\ref{eq:beatbnd})
to just
\[
 \frac{\Phi_1({\cal M}_k,t)}{\Phi_1({\cal M},t)} \geq 1+\epsilon .
\]

We first show that the required bound is met for the control column if
and only if the selected portfolio is made up entirely of rows from the
first $kn$ rows (i.e., those rows that contain a 1 in the control
column).  In particular, the adjustment row may not be included in the
portfolio.  The market average for the control column is simply
\[ \Phi_1({\cal M},t) = 
       \frac{kn}{\left\lceil(1+\epsilon)kn\right\rceil} .
\]
Obviously, when the portfolio ${\cal M}_k$ is made up entirely of
these rows, the portfolio average in the control column is 1, so we
can bound
\[  \frac{\Phi_1({\cal M}_k,t)}{\Phi_1({\cal M},t)} = 
       \frac{\left\lceil(1+\epsilon)kn\right\rceil}{kn} \geq
       1+\epsilon .
\]
On the other hand, when only $k-1$ or fewer of the portfolio rows
begin with a 1, then the portfolio average is at most $1-\frac{1}{k}$,
and so we can bound
\begin{eqnarray*}
  \frac{\Phi_1({\cal M}_k,t)}{\Phi_1({\cal M},t)} & \leq &
       \left(1-\frac{1}{k}\right)
        \frac{\left\lceil(1+\epsilon)kn\right\rceil}{kn}
 <        \left(1-\frac{1}{k}\right)
        \frac{(1+\epsilon)kn+1}{kn} \\
 & = &        \left(1-\frac{1}{k}\right)
          \left(1+\epsilon+\frac{1}{kn}\right)
 =
       1+\epsilon+\frac{1}{kn}-\frac{1}{k}-\frac{\epsilon}{k}-\frac{1}{k^2n} \\
 & = & 1+\epsilon+-\frac{n-1}{kn}-\frac{\epsilon}{k}-\frac{1}{k^2n}
 < 1+\epsilon .
\end{eqnarray*}
Therefore, the desired bound is met only if all $k$ selected rows
begin with a 1.

We next show that the desired bound for all other columns is met if
and only if at least one row must be selected that contains a non-zero
value.  If no such rows are selected, all selected rows contain 0 and
so the portfolio average is 0.  This clearly cannot meet our required
bound.  On the other hand, if even one row is included with a non-zero
value, then $\Phi_1({\cal M}_k,t)\geq \frac{1}{k}$, while the market
average for this column is clearly
$\frac{n}{\left\lceil(1+\epsilon)kn\right\rceil}$.  This leads to
\[  \frac{\Phi_1({\cal M}_k,t)}{\Phi_1({\cal M},t)} \geq
       \frac{1}{k}\,\frac{\left\lceil(1+\epsilon)kn\right\rceil}{n} \geq
       1+\epsilon ,
\]
and so the desired bound is met.  We note that in order to meet the
desired bound on all columns, the adjustment row must not be selected,
and therefore the non-zero value required in each column of the
portfolio must come from the indicator variables of the original set
cover problem.  Therefore, an acceptable portfolio exists if and
only if an acceptable set cover exists.
\QED

%

\end{document}